\documentclass{article}

\usepackage[preprint]{template}
\usepackage[utf8]{inputenc}
\usepackage[T1]{fontenc}
\usepackage{hyperref}
\usepackage{url}
\usepackage{array}
\usepackage{booktabs}
\usepackage{multirow}
\usepackage{amsfonts, amsthm, amsmath, amssymb, mathtools}
\usepackage{nicefrac}
\usepackage{microtype}
\usepackage{xcolor}
\usepackage{colortbl}
\definecolor{paperink}{HTML}{22333B}
\definecolor{paperaccent}{HTML}{5E503F}
\definecolor{winnerc}{HTML}{C6AC8F}
\definecolor{tiec}{HTML}{EAE0D5}
\definecolor{nac}{gray}{0.55}

\definecolor{gnInk}{HTML}{1A1A1A}
\definecolor{gnRule}{HTML}{4D4D4D}
\definecolor{gnSoft}{HTML}{F4F4F4}
\definecolor{gnH0}{HTML}{F1F1F1}
\definecolor{gnH1}{HTML}{C8C8C8}
\definecolor{gnH2}{HTML}{8E8E8E}
\definecolor{gnH3}{HTML}{4F4F4F}
\definecolor{gnH4}{HTML}{1F1F1F}
\definecolor{paperblue}{HTML}{4A90B8}
\definecolor{skyblue}{HTML}{87CEEB}
\usepackage{natbib}
\usepackage{graphicx}
\usepackage{algorithm}
\usepackage{algpseudocode}
\usepackage{float}
\usepackage{caption}
\usepackage{subcaption}
\usepackage{enumitem}
\usepackage{tikz}
\usetikzlibrary{positioning, arrows.meta, calc, fit, shapes.geometric, decorations.pathreplacing}
\usepackage{listings}

\definecolor{lstback}{HTML}{F4F4F4}
\definecolor{lstrule}{HTML}{D8D8D8}
\definecolor{lstkw}{HTML}{22333B}
\definecolor{lststr}{HTML}{5E503F}
\definecolor{lstcomm}{HTML}{8E8E8E}
\lstdefinestyle{gnpython}{
  language=Python,
  basicstyle=\ttfamily\footnotesize,
  keywordstyle=\color{lstkw}\bfseries,
  stringstyle=\color{lststr},
  commentstyle=\color{lstcomm}\itshape,
  showstringspaces=false,
  backgroundcolor=\color{lstback},
  frame=single,
  rulecolor=\color{lstrule},
  framesep=4pt,
  xleftmargin=2pt, xrightmargin=2pt,
  breaklines=true,
  breakatwhitespace=true,
  columns=fullflexible,
  upquote=true,
  keepspaces=true,
  literate={'}{{\textquotesingle}}1
}
\hypersetup{
  colorlinks=true,
  linkcolor=blue!80,
  citecolor=blue!80,
  urlcolor=blue!80
}

\title{%
  \textsc{GraphNetz}: Statistical Benchmarking of Graph Neural Networks with Paired Tests and Rank Aggregation 
}

\author{%
  Kleyton da Costa\thanks{Corresponding author.} \\
  Holistic AI \\
  London, UK \\
  \texttt{kleyton.vsc@gmail.com} \\
  \And
  Bernardo Modenesi \\
  University of Utah \\
  Salt Lake City, UT, USA \\
  \texttt{bernardo.modenesi@utah.edu} \\
}

\begin{document}
\maketitle

\begin{center}
\vspace{-0.6em}
\small
\textbf{Code:} \href{https://github.com/quant-sci/graphnetz}{github.com/quant-sci/graphnetz} \quad\textbf{$\bullet$}\quad
\textbf{Docs:} \href{https://graphnetz.readthedocs.io}{graphnetz.readthedocs.io}
\vspace{0.6em}
\end{center}

\begin{abstract}
Graph Neural Networks (GNNs) benchmarks often report single point estimates, even when performance differences are small relative to variation across random seeds, train/test splits, and datasets. Confidence intervals, paired comparisons, multiple-comparison correction, and rank-based aggregation are standard statistical tools, but they are rarely the default output of graph-learning benchmark suites. We introduce \textsc{GraphNetz}, a benchmarking framework whose default output is a structured statistical report rather than a raw accuracy table. \textsc{GraphNetz} currently includes 63 dataset loaders, four task types, and five canonical GNN architectures, while also supporting custom datasets and models. The framework standardizes multi-seed evaluation and automatically returns per-cell confidence intervals, Holm-corrected paired tests, and Friedman–Nemenyi critical-difference diagrams across tasks. In a cross-category benchmark over ten heterogeneous tasks, apparent rank differences among four canonical node-level encoders fall within a single Nemenyi clique, indicating that none is significantly better than the others at $\alpha = 0.05$. \textsc{GraphNetz} therefore provides researchers with a reproducible computational and statistical pipeline to benchmark new graph-learning methods against standard architectures, over different tasks and a wide set of applications, while reporting principled statistical evidence for benchmarking which accounts for seed uncertainty. This framework is set to serve the graph-learning community with a reproducible and honest model comparison ready to be added to papers.
\end{abstract}

\section{Introduction}
\label{sec:introduction}

Graph neural networks (GNNs) are now the default tool for problems as diverse as molecular property prediction, citation analysis, social network mining, anti-money laundering, and combinatorial optimization. However, a typical benchmarking paper in this area trains a few models on few datasets, reports a single test accuracy (plus standard deviation) per cell, and concludes that one model ``outperforms'' another. Headline differences are reported without confidence intervals, paired comparisons, multiple-comparison correction, or rank-based aggregation across datasets. This behavior can be viewed as a symptom of the recent reproducibility crises in artificial intelligence in general \citep{hutson2018artificial} and graph learning in specifics \citep{bechler2025position}. The result is a literature whose effect sizes are routinely smaller than the noise from which they are extracted, and can produce negative implications for the relevance of the field.

Two well-known re-evaluations established this concretely. \citet{shchur2018pitfalls} showed that the headline differences between GCN, GAT, and GraphSAGE on the Planetoid splits collapse once the seed and split are randomized; \citet{errica2020fair} reached the same conclusion for graph classification under a fair hyperparameter search, tracing much of the literature's reported ``progress'' to evaluation artifacts rather than architectural gains. Both works argue, in effect, that the right statistical apparatus must be embedded in the default pipeline, not bolted on by careful authors.

Using multi-seed Student's $t$ confidence intervals, paired $t$-tests (and their non-parametric Wilcoxon signed-rank alternatives) with Holm-Bonferroni correction~\citep{holm1979simple}, and rank-based aggregation with the Friedman omnibus and Nemenyi post-hoc, visualized as a critical-difference (CD) diagram~\citep{demsar2006statistical}, is half a century old and well understood. Nothing prevents a GNN practitioner from applying it; what is missing is a default surface that produces it without bespoke statistics code.

\paragraph{Where the existing tooling stops.}
PyTorch (PyG)~\citep{fey2019fast} is the main layer library for GNN research, exposing message-passing primitives and a broad collection of dataset loaders, and metrics. The library is deliberately \emph{architecture-first} and leave training loops, multi-seed orchestration, and statistical reporting to the user. The major benchmark suites: the Open Graph Benchmark~\citep{hu2020ogb}, Benchmarking-GNNs~\citep{dwivedi2020benchmarking}, and the recent GraphBench~\citep{stoll2025graphbench}\footnote{ The paper extends task coverage to node-level, edge-level, graph-level, and generative settings with out-of-distribution splits and unified hyperparameter tuning}. None of them makes per-cell CIs, multiple-comparison-corrected pairwise tests, or critical-difference diagrams the default return type of a benchmark run. Table~\ref{tab:positioning} summarizes this gap.

\paragraph{Our position.}
Statistical rigor in GNN benchmarking is both a tooling problem and a reproducibility one. If multi-seed, multiple-comparison-corrected, rank-aggregated reporting were the default surface of the dominant benchmarking framework, every benchmark paper would inherit it for free. We introduce \textsc{GraphNetz}, a benchmarking framework for the graph-learning community, whose default output is a structured statistical report rather than a raw accuracy table, utilizing a wide set of applications and state-of-the-art graph-learning models for a thorough and honest model comparison. This not only streamlines the benchmark process with a clear and reproducible pipeline, but also mitigates \textit{cherry-picking} concerns, comparing competing GNN methods to existing ones in a wide cast of scenarios, while accounting for seed uncertainty. To our knowledge, \textsc{GraphNetz} is the first GNN-focused benchmarking framework to make the full chain the default output of a benchmark run.

Our main contributions include:

\begin{itemize}
  \item A categorical dataset taxonomy that maps 63 loaders to ten research areas (combinatorial optimization, biology, social, knowledge, infrastructure, finance, computing, vision, physics, and security), with explicit notes on subtopics that lack canonical public benchmarks;
  \item A uniform benchmark protocol for five of the most used GNN architectures (GCN, GAT, GIN, GraphSAGE, Graph Transformer) across four task types (node classification, graph classification, graph regression, link prediction) so that differences in reported performance reflect modeling choices rather than evaluation idiosyncrasies;
  \item A statistically principled reporting layer,  mean $\pm$ Student's $t$ CI per cell, Holm--Bonferroni-corrected paired $t$-tests within tasks, and Friedman-Nemenyi critical-difference diagrams across tasks, formally defined in the Appendix~\ref{app:prelim} and instantiated as the default return type of every benchmark run;
  \item A cross-category empirical illustration of the full pipeline (Section~\ref{sec:experiment}) showing that with $N=10$ heterogeneous tasks under a uniform held-out-metric protocol, none of the four node-level encoders is significantly better than any other at $\alpha = 0.05$ — all four fall within a single Nemenyi clique ($CD_{0.05}=1.48$);
\end{itemize}

This paper is organized as follows:  Section~\ref{sec:design} describes the \textsc{GraphNetz} evaluation protocol and states it formally as Algorithm~\ref{alg:bench}. Section~\ref{sec:experiment} runs the cross-category benchmark and reports its results. Sections~\ref{sec:discussion} discuss the findings and limitations of the framework. The formal statistical machinery (Student's $t$ CI, paired-$t$ with Holm correction, Friedman--Nemenyi $CD$) is collected in Appendix~\ref{app:prelim}.

\section{The \textsc{GraphNetz} Evaluation Protocol}
\label{sec:design}

\textsc{GraphNetz} is organized around four conceptual layers: a categorized catalog of datasets, a set of model evaluations under a common protocol, training routines covering the standard task types, and the statistical reporting layer (formal definitions in Appendix~\ref{app:prelim}). Figure~\ref{fig:protocol} is the schematic. The Algorithm~\ref{alg:bench} formally states the same pipeline.

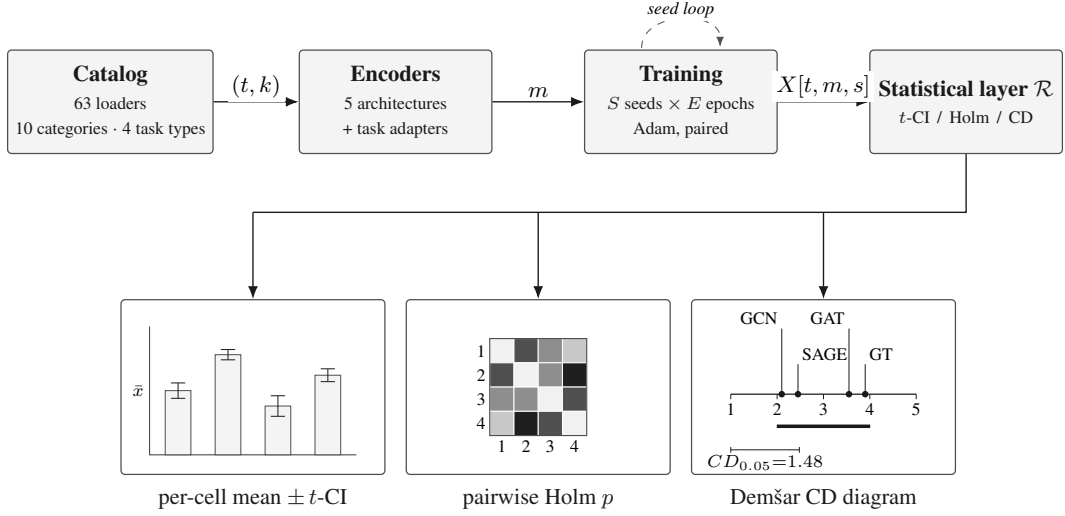
\begin{figure}[t]
\centering
\resizebox{\linewidth}{!}{%
\begin{tikzpicture}[
    font=\footnotesize,
    >=Latex,
    stage/.style={
        draw=gnRule, line width=0.5pt,
        rounded corners=1.6pt,
        fill=gnSoft,
        minimum width=2.7cm, minimum height=1.45cm,
        align=center, text=gnInk, inner sep=3pt,
    },
    flow/.style={-{Latex[length=1.8mm,width=1.4mm]}, line width=0.55pt, draw=gnInk},
    stem/.style={line width=0.55pt, draw=gnInk},
    flowlabel/.style={font=\footnotesize, fill=white, inner sep=1.5pt},
    panel/.style={
        draw=gnRule, line width=0.5pt,
        rounded corners=1.6pt, fill=white,
        minimum width=3.7cm, minimum height=2.45cm,
        anchor=center, inner sep=0pt,
    },
    paneltitle/.style={font=\footnotesize, text=gnInk},
]
% --- pipeline -----------------------------------------------------------
\node[stage] (cat)  at (0,0)  {{\bfseries Catalog}\\[1pt]{\scriptsize 63 loaders}\\{\scriptsize 10 categories $\cdot$ 4 task types}};
\node[stage] (enc)  at (4,0)  {{\bfseries Encoders}\\[1pt]{\scriptsize 5 architectures}\\{\scriptsize {+} task adapters}};
\node[stage] (tr)   at (8,0)  {{\bfseries Training}\\[1pt]{\scriptsize $S$ seeds $\times\, E$ epochs}\\{\scriptsize Adam, paired}};
\node[stage] (stat) at (12,0) {{\bfseries Statistical layer $\mathcal{R}$}\\[1pt]{\scriptsize $t$-CI \,/\, Holm \,/\, CD}};
\draw[flow] (cat) -- node[flowlabel,above] {$(t,k)$}    (enc);
\draw[flow] (enc) -- node[flowlabel,above] {$m$}        (tr);
\draw[flow] (tr)  -- node[flowlabel,above] {$X[t,m,s]$} (stat);
% seed loop
\draw[-{Latex[length=1.4mm,width=1.1mm]}, dashed, line width=0.4pt, draw=gnRule, shorten >=0.5pt]
  ($(tr.north)+(-0.55,0.02)$) .. controls ($(tr.north)+(-0.55,0.55)$) and ($(tr.north)+(0.55,0.55)$) .. ($(tr.north)+(0.55,0.02)$);
\node[font=\scriptsize\itshape, fill=white, inner sep=1pt] at ($(tr.north)+(0,0.55)$) {seed loop};
% --- manifold + panels --------------------------------------------------
\def\panelY{-4.0}
\node[panel] (p1) at (2,\panelY)  {};
\node[panel] (p2) at (6,\panelY)  {};
\node[panel] (p3) at (10,\panelY) {};
\def\manifoldY{-1.55}
\draw[stem] (stat.south) -- (12,\manifoldY) -- (2,\manifoldY);
\draw[flow] (2,\manifoldY)  -- (p1.north);
\draw[flow] (6,\manifoldY)  -- (p2.north);
\draw[flow] (10,\manifoldY) -- (p3.north);
\node[paneltitle, below=2pt of p1] {per-cell mean $\pm\, t$-CI};
\node[paneltitle, below=2pt of p2] {pairwise Holm $p$};
\node[paneltitle, below=2pt of p3] {Dem\v{s}ar CD diagram};
% --- panel 1: bars + t-CI ----------------------------------------------
\begin{scope}[shift={(p1.center)}]
    \def\xL{-1.45}\def\xR{1.45}\def\yB{-0.95}\def\yT{0.85}
    \pgfmathsetmacro{\plotH}{\yT-\yB}
    \draw[line width=0.45pt, draw=gnRule] (\xL,\yT) -- (\xL,\yB) -- (\xR,\yB);
    \node[font=\scriptsize, anchor=east, inner sep=2pt] at (\xL,-0.05) {$\bar{x}$};
    \foreach \i/\hf/\cif in {1/0.50/0.06, 2/0.78/0.04, 3/0.38/0.08, 4/0.62/0.05} {
        \pgfmathsetmacro{\xc}{-1.05 + (\i-1)*0.70}
        \pgfmathsetmacro{\barW}{0.36}
        \pgfmathsetmacro{\xLb}{\xc - \barW/2}\pgfmathsetmacro{\xRb}{\xc + \barW/2}
        \pgfmathsetmacro{\barTop}{\yB + \hf*\plotH}
        \pgfmathsetmacro{\ciLo}{\barTop - \cif*\plotH}\pgfmathsetmacro{\ciHi}{\barTop + \cif*\plotH}
        \draw[fill=gnSoft, draw=gnRule, line width=0.35pt] (\xLb,\yB) rectangle (\xRb,\barTop);
        \draw[line width=0.4pt, gnInk] (\xc,\ciLo) -- (\xc,\ciHi);
        \draw[line width=0.4pt, gnInk] (\xc-0.10,\ciLo) -- (\xc+0.10,\ciLo);
        \draw[line width=0.4pt, gnInk] (\xc-0.10,\ciHi) -- (\xc+0.10,\ciHi);
    }
\end{scope}
% --- panel 2: Holm heatmap ---------------------------------------------
\begin{scope}[shift={(p2.center)}]
    \def\cell{0.34}\def\matX{-0.68}\def\matY{-0.68}
    \foreach \i/\j/\c in {%
        1/1/gnH0,1/2/gnH3,1/3/gnH2,1/4/gnH1,
        2/1/gnH3,2/2/gnH0,2/3/gnH2,2/4/gnH4,
        3/1/gnH2,3/2/gnH2,3/3/gnH0,3/4/gnH3,
        4/1/gnH1,4/2/gnH4,4/3/gnH3,4/4/gnH0%
    } {
        \pgfmathsetmacro{\xx}{\matX + (\j-1)*\cell}
        \pgfmathsetmacro{\yy}{\matY + (4-\i)*\cell}
        \draw[fill=\c, draw=white, line width=0.4pt] (\xx,\yy) rectangle (\xx+\cell,\yy+\cell);
    }
    \draw[line width=0.45pt, draw=gnRule] (\matX,\matY) rectangle (\matX+4*\cell,\matY+4*\cell);
    \foreach \j in {1,2,3,4} {
        \pgfmathsetmacro{\xx}{\matX + (\j-1)*\cell + \cell/2}
        \node[font=\scriptsize, below, inner sep=1.5pt] at (\xx,\matY) {\j};
    }
    \foreach \i in {1,2,3,4} {
        \pgfmathsetmacro{\yy}{\matY + (4-\i)*\cell + \cell/2}
        \node[font=\scriptsize, left, inner sep=1.5pt] at (\matX,\yy) {\i};
    }
\end{scope}
% --- panel 3: Demsar CD diagram ----------------------------------------
\begin{scope}[shift={(p3.center)}]
    \def\axLx{-1.30}\def\axRx{1.30}\def\axY{-0.10}\def\rmin{1}\def\rmax{5}
    \pgfmathsetmacro{\axW}{\axRx-\axLx}
    \draw[line width=0.5pt, gnInk] (\axLx,\axY) -- (\axRx,\axY);
    \foreach \r in {1,2,3,4,5} {
        \pgfmathsetmacro{\xx}{\axLx + (\r-\rmin)/(\rmax-\rmin)*\axW}
        \draw[line width=0.4pt, gnInk] (\xx,\axY) -- (\xx,\axY-0.07);
        \node[font=\scriptsize, below, inner sep=1.5pt] at (\xx,\axY-0.07) {\r};
    }
    \node[font=\scriptsize, anchor=west, inner sep=1pt] at (\axRx+0.04,\axY) {};
    \foreach \r/\name/\dy/\lanchor in {%
        2.10/GCN/0.92/{south east},
        2.45/SAGE/0.42/{south west},
        3.55/GAT/0.92/{south east},
        3.90/GT/0.42/{south west}%
    } {
        \pgfmathsetmacro{\xx}{\axLx + (\r-\rmin)/(\rmax-\rmin)*\axW}
        \fill[gnInk] (\xx,\axY) circle (1.3pt);
        \draw[line width=0.4pt, gnInk] (\xx,\axY) -- (\xx,\axY+\dy);
        \node[font=\scriptsize, anchor=\lanchor, inner sep=1.5pt] at (\xx,\axY+\dy) {\name};
    }
    \pgfmathsetmacro{\xClL}{\axLx + (2.10-\rmin)/(\rmax-\rmin)*\axW - 0.07}
    \pgfmathsetmacro{\xClR}{\axLx + (3.90-\rmin)/(\rmax-\rmin)*\axW + 0.07}
    \draw[line width=1.4pt, gnInk] (\xClL,\axY-0.45) -- (\xClR,\axY-0.45);
    \pgfmathsetmacro{\cdLen}{1.48/(\rmax-\rmin)*\axW}
    \pgfmathsetmacro{\cdY}{\axY-0.78}
    \draw[line width=0.4pt, gnInk] (\axLx,\cdY) -- (\axLx+\cdLen,\cdY);
    \draw[line width=0.4pt, gnInk] (\axLx,\cdY-0.04) -- (\axLx,\cdY+0.04);
    \draw[line width=0.4pt, gnInk] (\axLx+\cdLen,\cdY-0.04) -- (\axLx+\cdLen,\cdY+0.04);
    \node[font=\scriptsize, anchor=north, inner sep=1.5pt]
        at (\axLx+\cdLen/2,\cdY-0.02) {$CD_{0.05}{=}1.48$};
\end{scope}
\end{tikzpicture}}
\caption{The \textsc{GraphNetz} evaluation protocol. \textbf{Top:} the processing pipeline
Catalog\,$\rightarrow$\,Encoders\,$\rightarrow$\,Training\,$\rightarrow$\,Statistical layer~$\mathcal{R}$;
arrows carry the data axes -- a (task $t$,\,data) pair, a model~$m$, and the seed-paired metric tensor
$X[t,m,s]$. The dashed arc on Training marks the multi-seed inner loop that builds the seed axis of~$X$.
\textbf{Bottom:} the three artifacts the framework emits by default -- per-cell mean with Student's
$t$ confidence interval, a Holm-corrected pairwise $p$-value matrix, and a Dem\v{s}ar
critical-difference diagram whose horizontal bar joins models within the Nemenyi threshold $CD$.}
\label{fig:protocol}
\end{figure}

\begin{algorithm}[t]
\caption{\textsc{GraphNetz} benchmark protocol.}
\label{alg:bench}
\begin{algorithmic}[1]
\Require category $C$, models $\mathcal{M}$, seeds $\mathcal{S}$,
  epochs $E$, level $\alpha$
\Ensure structured statistical report $\mathcal{R}$
\State $\mathcal{T} \gets \textsc{Tasks}(C)$
  \Comment{representative datasets and tasks for $C$}
\State $X \gets \emptyset$
  \Comment{tensor of per-seed final metrics indexed by (task, model, seed)}
\For{each task $t \in \mathcal{T}$}
  \For{each model $m \in \mathcal{M}$ \textbf{compatible with} $\textsc{Task}(t)$}
    \For{each seed $s \in \mathcal{S}$}
      \State seed all RNGs with $s$; instantiate $m$ with task-appropriate dimensions
      \State train $m$ on $t$ for $E$ epochs; record final metric $X[t,m,s]$
    \EndFor
  \EndFor
\EndFor
\State \textbf{Per-cell:} compute $\bar{x}_{t,m}$ and $h_{t,m}$ at level $\alpha$ from $X$
\State \textbf{Per-task pairwise:} for each $t$, compute paired-$t$ $p$-values for all
       $\binom{|\mathcal{M}_t|}{2}$ pairs and apply Holm--Bonferroni
\State \textbf{Across-task:} rank models per task, compute Friedman $\chi_F^2$
       and Nemenyi $CD_\alpha$
\State \Return $\mathcal{R} = (X, \bar{x}, h, \tilde{p}, \bar{r}, CD_\alpha)$
\end{algorithmic}
\end{algorithm}

\paragraph{Datasets.}
The catalog spans ten categories and is indexed by a \emph{category $\times$ task} taxonomy: each loader declares the tasks it can serve, so the same dataset can appear in more than one task (Cora, CiteSeer and PubMed, for example, are exposed as both node-classification and link-prediction tasks via PyG's
\texttt{RandomLinkSplit}). The taxonomy is materialized at module import time as \texttt{LOADER\_REGISTRY[category][task]}, and the curated benchmark mirrors the same shape as
\texttt{BENCHMARK\_TASKS[category][task]}. Where canonical public benchmarks exist (Planetoid~\citep{sen2008collective}, TUDataset~\citep{morris2020tudataset}, QM9, ZINC, ModelNet,
ShapeNet, Elliptic Bitcoin~\citep{weber2019elliptic}, MalNet~\citep{freitas2021malnet}, the Long-Range graph benchmark Peptides~\citep{dwivedi2022lrgb}, and the Heterophilous graph benchmark~\citep{platonov2023critical}), they are wrapped directly. Otherwise we draw from the Netzschleuder
catalog~\citep{peixoto2020netzschleuder} or generate synthetic instances (TSP, VRP, max-flow, bipartite matching, graph coloring, Ising lattice). Table~\ref{tab:categories} summarizes the loaders per category.

\paragraph{Models.}
The framework evaluates five architectures: GCN \citep{kipf2017semisupervised}, GAT \citep{velickovic2018graph}, GIN \citep{xu2019gin}, GraphSAGE \citep{hamilton2017graphsage}, and the
Graph Transformer \citep{shi2021masked}. Two task adapters, a graph-level pooling head over a node-level encoder and a dot-product link-prediction head, let the four node-level encoders enter every task in the benchmark (node classification, graph classification, graph regression, link prediction). GIN keeps its native graph-level pooling and is evaluated on the two graph-level. Task compatibility is checked before training, so incompatible (model, task) pairs are skipped automatically (Algorithm~\ref{alg:bench}, line 4). Deep Graph Infomax~\citep{velickovic2019dgi} is exposed as an optional
\emph{training utility} (\texttt{train\_dgi} + \texttt{DGIWrapper}) rather than a benchmark-task model: it is a self-supervised training objective whose ``metric'' is its own loss, so it cannot serve as a held-out test metric in the same row of the report as node-classification accuracy or link-prediction AUC.

\paragraph{Training.}
Five training routines cover the four tasks plus a self-supervised pre-training utility: node classification, graph classification, graph regression, link prediction (binary cross-entropy on a \texttt{RandomLinkSplit} with negative sampling, evaluated by ROC-AUC), and an optional DGI pre-training step that returns an unsupervised representation but is excluded from the default benchmark report because its loss is not a held-out metric. Each emits per-epoch metrics that feed the reporting layer
(Appendix~\ref{app:prelim}).

\paragraph{Programmatic interface.}
The algorithm~\ref{alg:bench} is exposed through a single entry point, \texttt{run\_benchmark}, that takes a category (or an explicit list of \texttt{Task} objects), a mapping from model names to model classes, a seed list, and an optional task filter; it returns a \texttt{BenchmarkReport} whose methods produce the artifacts of Figure~\ref{fig:protocol}. The same report object supports both the curated catalog and any user-defined task list (Appendix~\ref{app:templates}), so reviewers can reproduce the cells in this paper and extend the analysis with their own architectures or datasets without touching the internals of the framework.

\paragraph{Positioning.}
Table~\ref{tab:positioning} summarizes how \textsc{GraphNetz} differs from the closest related frameworks. The distinguishing axis is not architectural breadth or dataset count, but the presence of an out-of-the-box statistical reporting layer.

\begin{table}[t]
  \centering
  \caption{Positioning of \textsc{GraphNetz} against related GNN benchmarking frameworks. \emph{Coverage} counts built-it loaders and architectures (``open'' = user-supplied; ``$N$ + open'' = $N$ built-ins plus a one-line extension API). \emph{Statistical layer} flags whether the default report emits per-cell CIs ($t$ = Student's-$t$, \textit{std} = raw stdev), corrected pairwise tests, a Friedman--Nemenyi CD diagram, and a bootstrap CI option. \emph{Extensibility \& reproducibility} flag a one-line custom-dataset / custom-model API and automatic per-\emph{(task, model, seed)} RNG reseeding. \checkmark\ = first-class; -- = unsupported; \textit{partial} = available but not default. Counts reflect each framework's reference paper.}
  \label{tab:positioning}
  \footnotesize
  \setlength{\tabcolsep}{4pt}
  \renewcommand{\arraystretch}{1.10}
  \resizebox{\linewidth}{!}{%
  \begin{tabular}{@{}l cc cccc cc c@{}}
    \toprule
    & \multicolumn{2}{c}{Coverage}
    & \multicolumn{4}{c}{Statistical layer (default report)}
    & \multicolumn{2}{c}{Extensibility (1-line API)}
    & Repro. \\
    \cmidrule(lr){2-3} \cmidrule(lr){4-7} \cmidrule(lr){8-9} \cmidrule(lr){10-10}
    Framework
    & Data & Models
    & CI & Pairwise & CD & Bootstrap
    & Dataset & Model
    & Reseed \\
    \midrule
    PyTorch Geometric~\citep{fey2019fast}             & many         & many       & --        & --              & --            & --            & open       & open       & --            \\
    Open Graph Benchmark~\citep{hu2020ogb}            & 15+          & open       & std       & --              & --            & --            & open       & open       & --            \\
    Benchmarking-GNNs~\citep{dwivedi2020benchmarking} & 8            & 6          & std       & --              & --            & --            & open       & open       & partial       \\
    GraphBench~\citep{stoll2025graphbench}            & multi-domain & MPNN, GT   & std       & --              & --            & --            & open       & open       & partial       \\
    \textbf{\textsc{GraphNetz} (ours)}                         & \textbf{63 + open} & \textbf{5 + open} & \textbf{$t$} & \textbf{Holm} & \textbf{\checkmark} & \textbf{\checkmark} & \textbf{\checkmark} & \textbf{\checkmark} & \textbf{\checkmark} \\
    \bottomrule
  \end{tabular}}
\end{table}

\section{Cross-Category Experiment}
\label{sec:experiment}

\subsection{Protocol}
\label{sec:protocol}

We pick one representative dataset from each of the ten research categories and run the benchmark once per category with five architectures (GCN, GAT, GraphSAGE, Graph Transformer, GIN). The dispatcher trains each compatible (model, task, seed) combination in $S = 10$ seeds (0--9) and automatically skips incompatible pairs . Every (model, task, seed) combination trains independently with full-batch Adam optimizer \citep{kingma2014adam} and 64 hidden channels; the optimizer configuration follows the per-task defaults of the
framework's training routines: $\mathrm{lr}=10^{-2}$ with weight decay $5\times10^{-4}$ for node classification, $\mathrm{lr}=10^{-3}$ without weight decay for graph classification and graph regression, and $\mathrm{lr}=10^{-2}$ without weight decay for link prediction. The number of epochs $E$ is set per task to keep total runtime in a laptop-friendly budget (Cora, CiteSeer, PubMed: 100; MUTAG: 40; ZINC: 10; MNIST-superpixels: 4; FB15k-237: 20; Euroroad, Board-directors, Internet AS: 40--80; TSP-random and Ising-lattice: 60-80; 9/11 terrorists: 120). Random number generators are reset before each run, so the entire pipeline is deterministic given a fixed software stack.

The multi adapters described in Section~\ref{sec:design} register every node-level encoder for every task, so each category exposes \emph{at least four} models for comparison; categories whose natural task is graph classification expose five (the four node-level encoders wrapped with mean-pool plus the native GIN). The selected datasets and the corresponding tasks are listed in Table~\ref{tab:breadth}. The knowledge slot uses FB15k-237 with a relational \texttt{DistMult} decoder over the canonical \texttt{(train, valid, test)} edge split shipped by PyG's \texttt{RelLinkPredDataset}: relation types are kept and the score is $\langle z_h, r, z_t\rangle$. The AUC reported here is a baseline rather than the SOTA relational LP number, since the encoder uses fabricated 3-d log-degree node features and a single-pass DistMult head; replacing either with a stronger choice is straightforward but orthogonal to the methodological focus of this paper.

\subsection{Pre-registered statistical analysis}
\label{sec:analysis}

For each (task, model) pair we report mean $\pm$ 95\,\% Student's $t$ CI half-width (Eq.~2) across the ten seeds. Within each task we run \emph{two} pairwise tests on per-seed final metrics: a paired $t$-test (parametric) and a Wilcoxon signed-rank test (non-parametric); both are corrected with Holm--Bonferroni (Eq.~4) at $\alpha = 0.05$. The Wilcoxon variant follows the recommendation of \citet{benavoli2016ranks} and is included here as a robustness check on the parametric test's normality assumption. Across the ten executed categories we rank the four node-level encoders per task with the correct metric direction (higher accuracy and higher AUC are better, lower MAE is better), average the ranks, and compute the Nemenyi critical difference (Eq.~6) at $\alpha = 0.05$ with $k = 4$ and $N = 10$. The Friedman--Nemenyi step is rank-based and therefore inherits no parametric assumption from either pairwise test.

\subsection{Per-category results}

Figure~\ref{fig:breadth} renders one panel per category; each panel uses the natural metric for the underlying task (test accuracy for node classification on the canonical Planetoid split, validation accuracy on a per-run 80/20 in-script split for graph classification, validation MAE for graph regression, and test AUC for link prediction on a held-out edge split) and shows the mean $\pm$ 95\,\% CI across ten seeds. Table~\ref{tab:breadth} lists the same numbers in long format. Because every working category receives at least four architectures, the per-category bars admit the full within-task pairwise machinery (paired $t$ on seed-aligned final metrics with Holm--Bonferroni correction) without any change to the protocol.

\subsection{Across-category ranking}

Although the per-category metrics are not commensurable (test accuracy vs.\ test AUC vs.\ MAE), \emph{ranks} are. The Friedman statistic and Nemenyi post-hoc described in Appendix~\ref{app:prelim} yield the CD diagram of Figure~\ref{fig:breadth_cd}. The Friedman omnibus on the same rank table gives $\chi^2_3 = 2.04$ ($p = 0.56$), so we already fail to reject the global null of equal mean ranks at $\alpha = 0.05$
before applying any post-hoc. With $k = 4$ node-level encoders common to every category and $N = 10$ tasks, the Nemenyi critical difference is $CD_{0.05} = 1.48$ rank units. GraphSAGE has the lowest mean rank ($\bar{r} = 2.10$), followed by GCN ($2.40$), the Graph Transformer ($2.60$), and GAT ($2.90$). The maximum rank gap (between GraphSAGE and GAT) is $0.80$, which is below $CD$, so \emph{none} of the four architectures is significantly different from any other at $\alpha = 0.05$ across this benchmark: all four
fall in a single Nemenyi clique. This is a sharper version of the headline argument: under a uniform held-out test metric on heterogeneous tasks, headline differences between canonical GNN architectures dissolve at typical sample sizes.

\subsection{Sensitivity to the pairwise test}
\label{sec:wilcoxon}

The two pairwise tests of Section~\ref{sec:analysis} agree on every within-task comparison at the seed count used here. Table~\ref{tab:pairwise_compare} reports the side-by-side adjusted $p$-values on the social/Planetoid demo (Cora, CiteSeer, PubMed, $3\times\binom{4}{2} = 18$ comparisons): both the paired $t$-test and the Wilcoxon signed-rank flag \emph{ten} comparisons as significant at $\alpha = 0.05$ after Holm--Bonferroni correction, and disagree on none. The agreement is not coincidence: it is a structural consequence of running the demo at $S=10$ rather than at smaller seed counts where the exact Wilcoxon null is too coarse to clear Holm correction.

\paragraph{Why Wilcoxon and paired $t$ agree at $S = 10$.} 
With $S = 10$ paired observations and a two-sided exact null, the smallest achievable Wilcoxon $p$-value is $2 \cdot 2^{-S} = 1/512 \approx 0.00195$, attained when all ten paired differences share the same sign. Six pairwise comparisons per task multiply through Holm to $0.00195 \times 6 \approx 0.0117$, which is well below the $\alpha = 0.05$ threshold. The paired $t$-test, by contrast, has a continuous null and can drive $p$-values arbitrarily low when the seeds agree strongly. At $S=10$ both tests have sufficient power to detect the systematic differences visible in the Planetoid demo; the agreement on all $18$ comparisons is a reassurance that the parametric assumption is not misleading here. At smaller seed counts ($S=5$ gives a smallest adjusted Wilcoxon $p \approx 0.375$) the non-parametric test would be underpowered and the two methods would diverge, which is why the framework recommends Wilcoxon as the small-sample default while reporting both.

\paragraph{Recommendation.}
We follow \citet{benavoli2016ranks} in treating the Wilcoxon signed-rank test as the small-sample-friendly default for classifier comparisons, while reporting the paired $t$-test in parallel as the more powerful (but more parametric) alternative. The framework's default report includes both (\texttt{report.pairwise(method="t")} and \texttt{report.pairwise(method="wilcoxon")}) so the reader can inspect the agreement directly; pairs where the two tests agree deserve more confidence than pairs where they disagree.

\begin{table}[H]
  \centering
  \footnotesize
  \caption{Per-task pairwise tests on the social/Planetoid sweep
    (Cora, CiteSeer, PubMed; $S=10$ seeds, $3\times\binom{4}{2}=18$
    comparisons). $\Delta\mu$ is the mean accuracy difference
    (positive = first model better). $p_{\text{Holm}}^{\,t}$ and
    $p_{\text{Holm}}^{\,W}$ are Holm-adjusted paired-$t$ and
    Wilcoxon signed-rank $p$-values, with \colorbox{winnerc}{\strut\textbf{khaki}}
    marking significance at $\alpha = 0.05$. \colorbox{tiec}{\strut almond}
    marks non-significant pairs.}
  \label{tab:pairwise_compare}
  %\resizebox{\linewidth}{!}{\input{tables/pairwise_compare.tex}
  \begin{tabular}{l l c c c c}
\toprule
Task & Comparison & $\Delta\mu$ & $p_{\text{Holm}}^{\,t}$ & $p_{\text{Holm}}^{\,W}$ & Agree \\
\midrule
\multirow{6}{*}{\textbf{Cora}} & GAT \textsc{vs.} GCN & $-0.018$ & \cellcolor{winnerc}{\color{paperink}\textbf{0.000523}} & \cellcolor{winnerc}{\color{paperink}\textbf{0.0117}} & {\color{paperink}\checkmark} \\
 & GAT \textsc{vs.} GraphSAGE & $-0.006$ & \cellcolor{tiec}{\color{paperaccent}0.341} & \cellcolor{tiec}{\color{paperaccent}0.547} & {\color{paperink}\checkmark} \\
 & GAT \textsc{vs.} GraphTransformer & $-0.001$ & \cellcolor{tiec}{\color{paperaccent}0.778} & \cellcolor{tiec}{\color{paperaccent}0.713} & {\color{paperink}\checkmark} \\
 & GCN \textsc{vs.} GraphSAGE & $+0.012$ & \cellcolor{winnerc}{\color{paperink}\textbf{3.58e-05}} & \cellcolor{winnerc}{\color{paperink}\textbf{0.0117}} & {\color{paperink}\checkmark} \\
 & GCN \textsc{vs.} GraphTransformer & $+0.017$ & \cellcolor{winnerc}{\color{paperink}\textbf{0.000149}} & \cellcolor{winnerc}{\color{paperink}\textbf{0.0117}} & {\color{paperink}\checkmark} \\
 & GraphSAGE \textsc{vs.} GraphTransformer & $+0.005$ & \cellcolor{tiec}{\color{paperaccent}0.176} & \cellcolor{tiec}{\color{paperaccent}0.264} & {\color{paperink}\checkmark} \\
\midrule
\multirow{6}{*}{\textbf{CiteSeer}} & GAT \textsc{vs.} GCN & $-0.043$ & \cellcolor{winnerc}{\color{paperink}\textbf{0.000139}} & \cellcolor{winnerc}{\color{paperink}\textbf{0.0117}} & {\color{paperink}\checkmark} \\
 & GAT \textsc{vs.} GraphSAGE & $-0.050$ & \cellcolor{winnerc}{\color{paperink}\textbf{0.000274}} & \cellcolor{winnerc}{\color{paperink}\textbf{0.0117}} & {\color{paperink}\checkmark} \\
 & GAT \textsc{vs.} GraphTransformer & $-0.060$ & \cellcolor{winnerc}{\color{paperink}\textbf{9.06e-06}} & \cellcolor{winnerc}{\color{paperink}\textbf{0.0117}} & {\color{paperink}\checkmark} \\
 & GCN \textsc{vs.} GraphSAGE & $-0.008$ & \cellcolor{tiec}{\color{paperaccent}0.245} & \cellcolor{tiec}{\color{paperaccent}0.164} & {\color{paperink}\checkmark} \\
 & GCN \textsc{vs.} GraphTransformer & $-0.017$ & \cellcolor{winnerc}{\color{paperink}\textbf{5.12e-05}} & \cellcolor{winnerc}{\color{paperink}\textbf{0.0117}} & {\color{paperink}\checkmark} \\
 & GraphSAGE \textsc{vs.} GraphTransformer & $-0.009$ & \cellcolor{tiec}{\color{paperaccent}0.245} & \cellcolor{tiec}{\color{paperaccent}0.164} & {\color{paperink}\checkmark} \\
\midrule
\multirow{6}{*}{\textbf{PubMed}} & GAT \textsc{vs.} GCN & $-0.027$ & \cellcolor{winnerc}{\color{paperink}\textbf{0.000108}} & \cellcolor{winnerc}{\color{paperink}\textbf{0.0117}} & {\color{paperink}\checkmark} \\
 & GAT \textsc{vs.} GraphSAGE & $-0.004$ & \cellcolor{tiec}{\color{paperaccent}0.444} & \cellcolor{tiec}{\color{paperaccent}0.5} & {\color{paperink}\checkmark} \\
 & GAT \textsc{vs.} GraphTransformer & $+0.003$ & \cellcolor{tiec}{\color{paperaccent}0.48} & \cellcolor{tiec}{\color{paperaccent}0.574} & {\color{paperink}\checkmark} \\
 & GCN \textsc{vs.} GraphSAGE & $+0.023$ & \cellcolor{winnerc}{\color{paperink}\textbf{1.93e-06}} & \cellcolor{winnerc}{\color{paperink}\textbf{0.0117}} & {\color{paperink}\checkmark} \\
 & GCN \textsc{vs.} GraphTransformer & $+0.030$ & \cellcolor{winnerc}{\color{paperink}\textbf{7.14e-07}} & \cellcolor{winnerc}{\color{paperink}\textbf{0.0117}} & {\color{paperink}\checkmark} \\
 & GraphSAGE \textsc{vs.} GraphTransformer & $+0.007$ & \cellcolor{tiec}{\color{paperaccent}0.0818} & \cellcolor{tiec}{\color{paperaccent}0.0703} & {\color{paperink}\checkmark} \\
\bottomrule
\end{tabular}

\end{table}

\subsection{Scope of the experiment}

\begin{figure}[H]
  \centering
  \includegraphics[width=\linewidth]{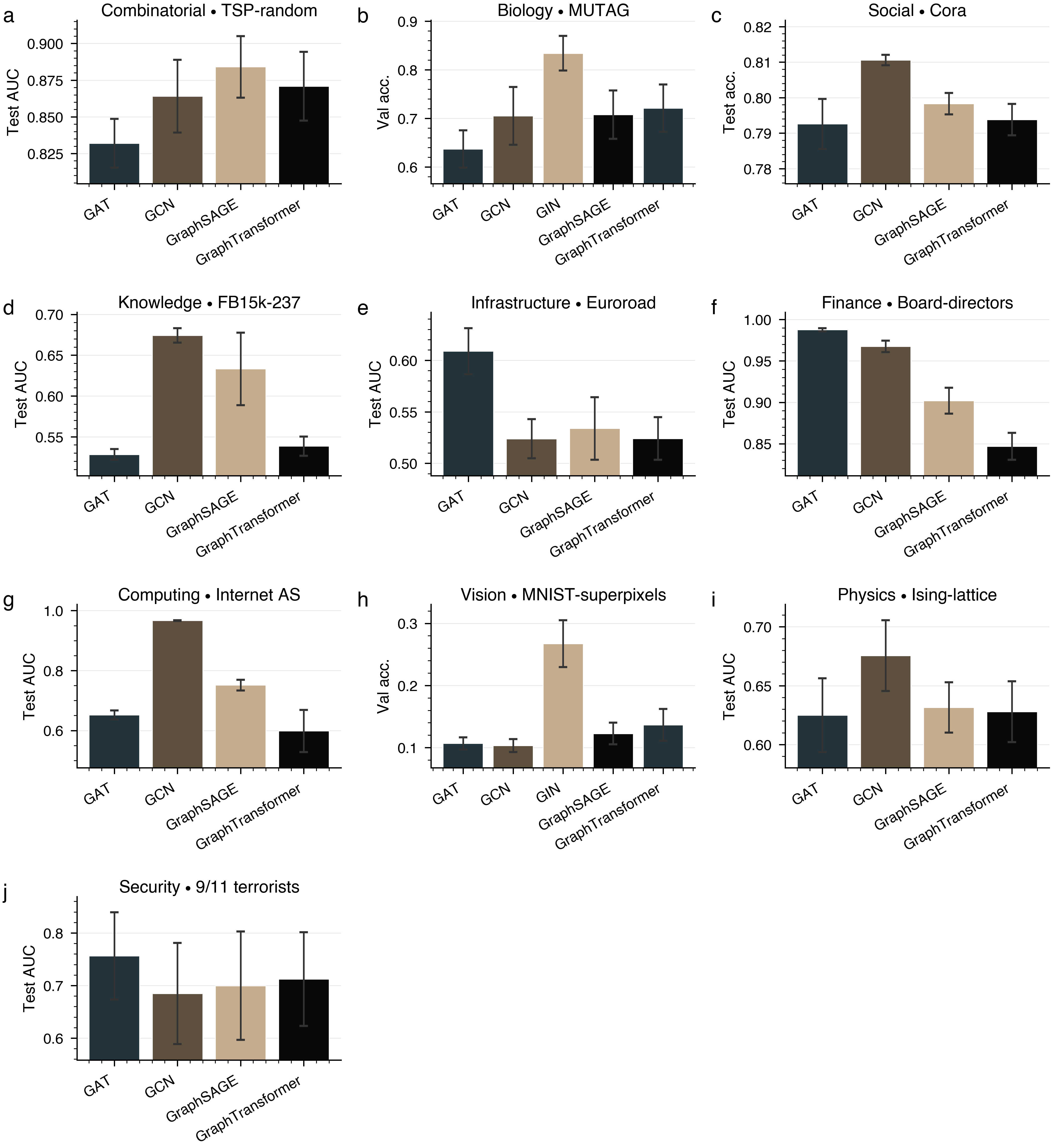}
  \caption{Cross-category dashboard: one representative dataset per research area, every compatible architecture trained across the same ten seeds. %Each panel uses the natural metric for the underlying task: test accuracy on the canonical Planetoid train/val/test split for node classification, validation accuracy on a per-run 80/20 in-script split for graph classification (TUDataset and MNIST-superpixels ship no canonical test split here), validation MAE for graph regression, and test AUC for link prediction on a held-out edge split. Each bar is the seed mean with a 95\,\% Student's $t$ confidence interval. All ten research categories are represented; the FB15k-237 panel uses the relational DistMult head over the canonical \texttt{RelLinkPredDataset} split described in Section~\ref{sec:protocol}.
  }
  \label{fig:breadth}
\end{figure}

The cross-category experiment \emph{is} a multi-architecture, multi-category benchmark: every working category receives at least four architectures with paired-by-seed confidence intervals on a uniform metric, and the four-encoder slice common to all categories admits a proper across-task CD diagram. It is \emph{not} a single uniform-metric leaderboard, and the per-category bar in Figure~\ref{fig:breadth} should not be compared against the bar in a neighboring category. Homogeneous head-to-head comparisons must
restrict attention to a single task.

\begin{table}[H]
  \centering
  \caption{Cross-category benchmark results. Each row is one
    representative dataset; each cell is the seed mean
    $x_{\pm h}$ with $h$ the 95\,\% Student's $t$ CI half-width
    across ten seeds. The metric direction column gives the
    ``best'' direction ($\uparrow$ higher / $\downarrow$ lower).
    \colorbox{winnerc}{\strut\textbf{khaki}} marks the best
    architecture per row; \colorbox{tiec}{\strut almond} marks
    architectures whose 95\,\% CI overlaps that of the winner
    (statistically indistinguishable at this sample size);
    {\color{nac}\textit{n/a}} marks architectures that are
    incompatible with the task.}
  \label{tab:breadth}
  \scriptsize
  \resizebox{\linewidth}{!}{\begin{tabular}{lllcccccc}
\toprule
Category & Dataset & Kind & Metric & GCN & GAT & SAGE & GT & GIN \\
\midrule
Combinatorial & TSP-random & link\_pred & AUC\,$\uparrow$ & \cellcolor{tiec}{\color{paperaccent}$0.864_{\pm 0.025}$} & $0.832_{\pm 0.017}$ & \cellcolor{winnerc}{\color{paperink}\textbf{$0.884_{\pm 0.021}$}} & \cellcolor{tiec}{\color{paperaccent}$0.871_{\pm 0.023}$} & {\color{nac}\textit{n/a}} \\
Biology & MUTAG & graph\_cls & Acc.\,$\uparrow$ & $0.705_{\pm 0.059}$ & $0.637_{\pm 0.038}$ & $0.708_{\pm 0.050}$ & $0.721_{\pm 0.049}$ & \cellcolor{winnerc}{\color{paperink}\textbf{$0.834_{\pm 0.036}$}} \\
Social & Cora & node\_cls & Acc.\,$\uparrow$ & \cellcolor{winnerc}{\color{paperink}\textbf{$0.811_{\pm 0.001}$}} & $0.793_{\pm 0.007}$ & $0.798_{\pm 0.003}$ & $0.794_{\pm 0.004}$ & {\color{nac}\textit{n/a}} \\
Knowledge & FB15k-237 & link\_pred & AUC\,$\uparrow$ & \cellcolor{winnerc}{\color{paperink}\textbf{$0.674_{\pm 0.009}$}} & $0.528_{\pm 0.007}$ & \cellcolor{tiec}{\color{paperaccent}$0.633_{\pm 0.044}$} & $0.538_{\pm 0.012}$ & {\color{nac}\textit{n/a}} \\
Infrastructure & Euroroad & link\_pred & AUC\,$\uparrow$ & $0.524_{\pm 0.019}$ & \cellcolor{winnerc}{\color{paperink}\textbf{$0.609_{\pm 0.022}$}} & $0.534_{\pm 0.030}$ & $0.524_{\pm 0.021}$ & {\color{nac}\textit{n/a}} \\
Finance & Board-directors & link\_pred & AUC\,$\uparrow$ & $0.968_{\pm 0.007}$ & \cellcolor{winnerc}{\color{paperink}\textbf{$0.988_{\pm 0.002}$}} & $0.902_{\pm 0.016}$ & $0.847_{\pm 0.016}$ & {\color{nac}\textit{n/a}} \\
Computing & Internet AS & link\_pred & AUC\,$\uparrow$ & \cellcolor{winnerc}{\color{paperink}\textbf{$0.967_{\pm 0.001}$}} & $0.652_{\pm 0.015}$ & $0.751_{\pm 0.018}$ & $0.598_{\pm 0.070}$ & {\color{nac}\textit{n/a}} \\
Vision & MNIST-superpixels & graph\_cls & Acc.\,$\uparrow$ & $0.103_{\pm 0.010}$ & $0.107_{\pm 0.010}$ & $0.123_{\pm 0.017}$ & $0.137_{\pm 0.026}$ & \cellcolor{winnerc}{\color{paperink}\textbf{$0.267_{\pm 0.038}$}} \\
Physics & Ising-lattice & link\_pred & AUC\,$\uparrow$ & \cellcolor{winnerc}{\color{paperink}\textbf{$0.676_{\pm 0.030}$}} & \cellcolor{tiec}{\color{paperaccent}$0.625_{\pm 0.031}$} & \cellcolor{tiec}{\color{paperaccent}$0.632_{\pm 0.021}$} & \cellcolor{tiec}{\color{paperaccent}$0.628_{\pm 0.026}$} & {\color{nac}\textit{n/a}} \\
Security & 9/11 terrorists & link\_pred & AUC\,$\uparrow$ & \cellcolor{tiec}{\color{paperaccent}$0.685_{\pm 0.096}$} & \cellcolor{winnerc}{\color{paperink}\textbf{$0.756_{\pm 0.083}$}} & \cellcolor{tiec}{\color{paperaccent}$0.700_{\pm 0.103}$} & \cellcolor{tiec}{\color{paperaccent}$0.712_{\pm 0.089}$} & {\color{nac}\textit{n/a}} \\
\bottomrule
\end{tabular}
}
\end{table}

\begin{figure}[H]
  \centering
  \includegraphics[width=0.95\linewidth]{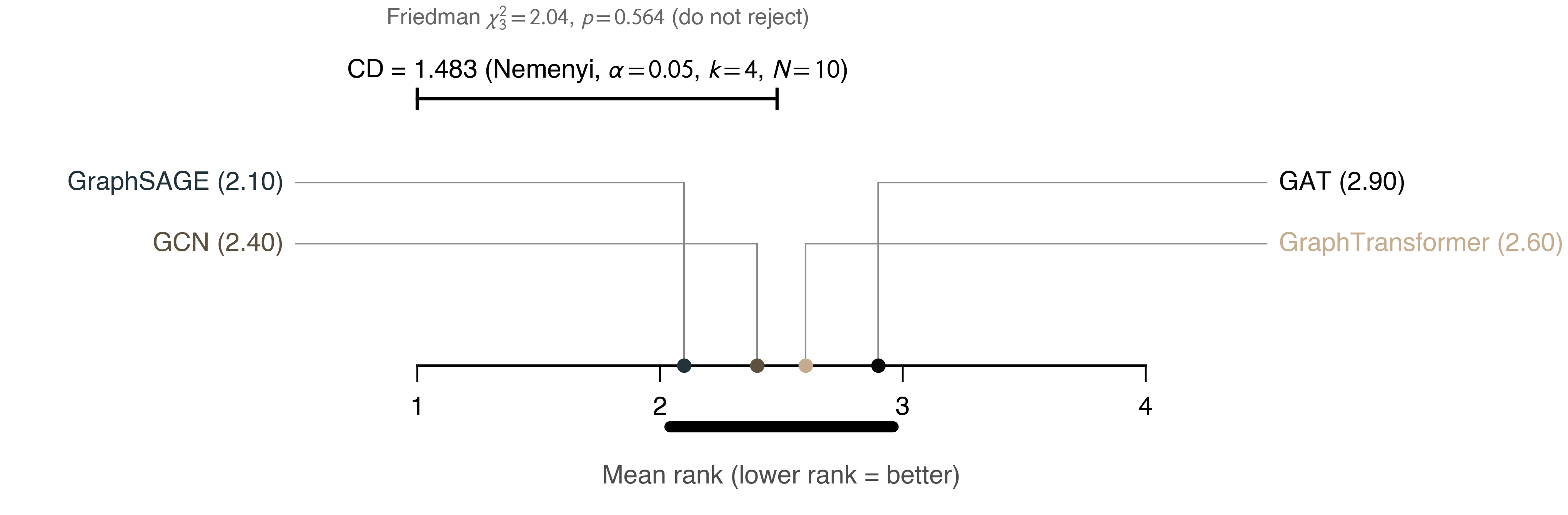}
  \caption{
  Cross-category Demšar critical-difference (CD) diagram at $\alpha = 0.05$ across $N = 10$ tasks. The diagram is restricted to the four node-level encoders (GCN, GAT, GraphSAGE, Graph Transformer) that are present in every category. % Per-task ranks use the correct direction (higher accuracy / AUC, lower MAE) and are averaged across categories. With $N = 10$ datasets the Nemenyi $CD$ is $1.48$ rank units. GraphSAGE has the lowest mean rank ($\bar{r}=2.10$), followed by GCN ($2.40$), the Graph Transformer ($2.60$), and GAT ($2.90$); the largest rank gap is $0.80 < CD$, so all four architectures fall in a single Nemenyi clique and none is significantly better than any other at this sample size.
  }
  \label{fig:breadth_cd}
\end{figure}

\begin{table}[H]
  \centering
  \caption{Dataset loaders per research category entries. Note: {\color{paperink}\textbf{NC}}~node classification, {\color{paperink}\textbf{GC}}~graph classification, {\color{paperink}\textbf{GR}}~graph regression, {\color{paperink}\textbf{LP}}~link prediction.}
  \label{tab:categories}
  \small
  \begin{tabular}{l r l >{\raggedright\arraybackslash}p{0.62\linewidth}}
\toprule
Category & \# & Kind & Loaders \\
\midrule
\multirow{1}{*}{Combinatorial} & \multirow{1}{*}{6} & {\color{paperink}\textbf{LP}} & bipartite matching, TSP, VRP, coloring, max-cut, max-flow \\
\midrule
\multirow{3}{*}{Biology} & \multirow{3}{*}{12} & {\color{paperink}\textbf{GC}} & MUTAG, PROTEINS, ENZYMES, Peptides-func, ogbg-molhiv, ogbg-molpcba \\
 &  & {\color{paperink}\textbf{GR}} & Peptides-struct \\
 &  & {\color{paperink}\textbf{LP}} & C.\,elegans, Budapest connectome, hospital contacts, high-school contacts, PPI \\
\midrule
\multirow{2}{*}{Social} & \multirow{2}{*}{16} & {\color{paperink}\textbf{NC}} & Cora, CiteSeer, PubMed, WikiCS, Roman-empire, Amazon-ratings, Minesweeper, Tolokers, Questions, ogbn-arxiv \\
 &  & {\color{paperink}\textbf{LP}} & Cora, CiteSeer, PubMed, MovieLens-100k, Karate, Facebook friends, DBLP coauthor, DNC emails, ogbl-collab \\
\midrule
\multirow{1}{*}{Knowledge} & \multirow{1}{*}{3} & {\color{paperink}\textbf{LP}} & FB15k-237, WordNet18-RR, WordNet (Netz) \\
\midrule
\multirow{1}{*}{Infrastructure} & \multirow{1}{*}{6} & {\color{paperink}\textbf{LP}} & power grid, EuroRoad, US roads, EU airlines, London transport, urban streets \\
\midrule
\multirow{2}{*}{Finance} & \multirow{2}{*}{5} & {\color{paperink}\textbf{NC}} & Elliptic Bitcoin, ogbn-products \\
 &  & {\color{paperink}\textbf{LP}} & product space, board of directors, US patents \\
\midrule
\multirow{1}{*}{Computing} & \multirow{1}{*}{4} & {\color{paperink}\textbf{LP}} & Internet AS, Internet topology, AS-Skitter, route views \\
\midrule
\multirow{2}{*}{Vision} & \multirow{2}{*}{5} & {\color{paperink}\textbf{NC}} & ShapeNet \\
 &  & {\color{paperink}\textbf{GC}} & MNIST superpixels, CIFAR-10 superpixels, ModelNet10, ModelNet40 \\
\midrule
\multirow{2}{*}{Physics} & \multirow{2}{*}{3} & {\color{paperink}\textbf{GR}} & QM9, ZINC \\
 &  & {\color{paperink}\textbf{LP}} & Ising lattice \\
\midrule
\multirow{2}{*}{Security} & \multirow{2}{*}{3} & {\color{paperink}\textbf{GC}} & MalNet-Tiny \\
 &  & {\color{paperink}\textbf{LP}} & 9/11 terrorists, train terrorists \\
\midrule
\textbf{Total} & \textbf{63} & & \\
\bottomrule
\end{tabular}

\end{table}

\section{Discussion and Limitations}
\label{sec:discussion}

The cross-category result is a sharper version of the prior empirical finding that headline differences among GNN architectures often disappear once seed-level variance and multiple comparisons are properly accounted for~\citep{shchur2018pitfalls,errica2020fair}. The CD diagram makes the residual signal precise: at $N = 10$ heterogeneous tasks under uniform held-out test metrics, the largest mean-rank gap among the four node-level encoders is $0.80$, below the Nemenyi threshold $CD_{0.05} = 1.48$, so none of the architectures is significantly different from any other at $\alpha = 0.05$. The methodological implication is that the default report, per-cell CI, per-task pairwise Holm, across-task Friedman-Nemenyi, is precisely calibrated to the kind of evidence the field actually has at typical sample sizes.

\paragraph{Number of datasets.}
Friedman--Nemenyi power scales with the number of datasets $N$: $CD_\alpha$ shrinks as $\sqrt{1/N}$. With $N = 10$ the diagram is informative about rank \emph{ordering} but unable to certify the small mean-rank gaps observed here as significant; lifting more of the catalogue (e.g.\ the heterophilic node-classification suite, the LRGB Peptides graphs, additional Netzschleuder snapshots) into the sweep would tighten $CD$ proportionally and is the right next experiment.

\paragraph{Fixed-epoch evaluation.}
The benchmark currently reports the test metric at a fixed final epoch. Checkpoint selection on a held-out validation set, early stopping, and best-of-window aggregation are natural extensions to the reporting layer that do not change the statistical pipeline of Appendix~\ref{app:prelim}.

\paragraph{Fixed hyperparameters.}
We hold optimisation hyperparameters fixed across categories. \citet{errica2020fair} argue that benchmarks should also be honest over the hyperparameter grid; extending Algorithm~\ref{alg:bench} with an inner search loop is straightforward but multiplies compute.

\paragraph{Coverage gaps in the taxonomy.}
Several subtopics still lack canonical public benchmarks: scheduling and bin packing in combinatorics; patient-disease-treatment knowledge graphs in biology; inter-bank exposure in finance; program control-flow, dataflow, build-dependency, and EDA graphs in computing; scene, skeleton, and SLAM pose graphs in vision; Feynman diagrams and reaction networks in physics; attack and threat-intelligence graphs in security. We document these gaps explicitly so that downstream users do not assume coverage that is absent.

\paragraph{Parametric assumptions.}
The paired $t$-test assumes approximate normality of per-seed differences. With $S = 10$ seeds the test is robust under mild departures from normality. The framework exposes two alternatives on the same paired-by-seed scaffolding: a percentile-bootstrap CI for skewed metrics (Appendix~\ref{app:bootstrap}) and a non-parametric Wilcoxon signed-rank pairwise test (\texttt{report.pairwise(method="wilcoxon")}). The latter is the small-sample-friendly default recommended by \citet{benavoli2016ranks} for classifier comparisons; pass
\texttt{method="wilcoxon"} or set \texttt{report.pairwise\_method} once and the Holm correction, the LaTeX exporter, and the pairwise significance plots all switch over.

\section{Conclusion}
\label{sec:conclusion}

GNN benchmarking sits at contrasting methodological grounds: the field has the statistical apparatus it needs, but the apparatus is not part of the default reporting pipeline. \textsc{GraphNetz} packages a category $\times$ task dataset taxonomy, a uniform model-evaluation protocol across four tasks categories, and a statistically structured reporting layer behind a benchmark whose default output is a structured statistical report. A single cross category run across ten heterogeneous tasks produces seed-paired confidence intervals per category and a Friedman-Nemenyi CD diagram between categories in which all four node-level encoders fall within a single indistinguishable clique at $\alpha = 0.05$, a calibrated picture
of what the data actually support.

\paragraph{Reproducibility.}
The benchmark protocol is fully specified by Algorithm~\ref{alg:bench} and the hyperparameters in Section~\ref{sec:protocol}. All random number generators are seeded, so given a fixed software stack, the pipeline is deterministic.

\paragraph{Data and code availability.}
All datasets used in Section~\ref{sec:experiment} are released under their original public licenses; loaders for them are part of the \textsc{GraphNetz} catalog. Synthetic categories (combinatorial, physics) are generated deterministically from the protocol seeds.

The natural research and implementation trajectory for \textsc{GraphNetz} involves expanding into graph foundation models and relational datasets, while prioritizing graph learning applications in real-world and scalable research environments.

% \paragraph{Author contributions.}
% K.dC.\ designed the framework, conducted the experiments, and wrote the manuscript. B.M.\ contributed to the statistical methodology and revised the manuscript. Both authors approved the final version.

% \begin{ack}
% We thank the maintainers of PyTorch Geometric, the Open Graph
% Benchmark, and the Netzschleuder catalogue, on which this work
% depends.
% \end{ack}

\bibliographystyle{abbrvnat}
\bibliography{references}

\appendix

\section{Statistical Overview}
\label{app:prelim}

We fix notation for the three layers of the statistical report produced by Algorithm~\ref{alg:bench}.

\paragraph{Per-cell estimation.}
For a fixed task and model, let $\{x_s\}_{s=1}^{S}$ be the per-seed final test metrics from $S$ independent training runs. The unbiased estimators of the population mean and variance are

\begin{equation}
  \bar{x} = \tfrac{1}{S}\sum_{s=1}^{S} x_s,
  \qquad
  \hat{\sigma}^2 = \tfrac{1}{S-1}\sum_{s=1}^{S}(x_s - \bar{x})^2.
\end{equation}
The Student's $t$ $(1-\alpha)$ confidence interval for the mean is
$\bar{x} \pm h$ with half-width
\begin{equation}
  h = t_{S-1, 1-\alpha/2}\,\hat{\sigma}/\sqrt{S},
\end{equation}
where $t_{S-1, 1-\alpha/2}$ is the $1-\alpha/2$ quantile of the
$t$-distribution with $S-1$ degrees of freedom. We report
$\bar{x} \pm h$ throughout.

\paragraph{Per-task pairwise tests.}
For a fixed task, let $x_s^{(i)}$ and $x_s^{(j)}$ denote the per-seed final metrics for models $i$ and $j$, and let $d_s = x_s^{(i)} - x_s^{(j)}$. Pairing by seed cancels seed-level variance. The paired-$t$ statistic
\begin{equation}
  T_{ij} = \frac{\bar{d}}{\hat{\sigma}_d/\sqrt{S}},
\end{equation}
with $\bar{d}$ and $\hat{\sigma}_d$ the sample mean and standard deviation of $d_s$, is referred to a $t_{S-1}$ null. The $\binom{k}{2}$ pairwise tests \emph{within} a task are corrected with the Holm-Bonferroni step-down procedure~\citep{holm1979simple}: order raw $p$-values $p_{(1)} \le \ldots \le p_{(m)}$ with $m = \binom{k}{2}$ and define adjusted values
\begin{equation}
  \tilde{p}_{(r)} = \min\!\Bigl(1,\ \max_{r' \le r}\bigl[(m - r' + 1)\,p_{(r')}\bigr]\Bigr).
\end{equation}
Holm controls the family-wise error rate at $\alpha$ and is uniformly more powerful than plain Bonferroni. We additionally report the paired effect size as Cohen's $d_z = \bar{d}/\hat{\sigma}_d$, which is invariant to $S$ and complements the $p$-value with a magnitude.

\paragraph{Wilcoxon signed-rank alternative.}
The paired $t$-test relies on approximate normality of $d_s$, which at small $S$ is fragile. The framework exposes a non-parametric alternative on the same paired-by-seed scaffolding: the Wilcoxon signed-rank test on $\{d_s\}$. Let $R_s = \mathrm{rank}(|d_s|)$ be the rank of each absolute paired difference (omitting ties at zero, following \texttt{zero\_method="wilcox"}); the signed-rank statistic
\begin{equation}
  W \;=\; \sum_{s\,:\,d_s > 0} R_s
  \;-\; \sum_{s\,:\,d_s < 0} R_s
\end{equation}
is referred to its exact null at small $S$ rather than to a normal approximation. The resulting $p$-values feed the same Holm adjustment of Eq.~4. \citet{benavoli2016ranks} argue this is the small-sample-friendly default for classifier comparisons; we expose it through \texttt{report.pairwise(method="wilcoxon")} (or the sticky \texttt{report.pairwise\_method} attribute), and the same flag flows into \texttt{pairwise\_to\_latex} and the pairwise significance plots.

\paragraph{Across-task aggregation.}
For benchmarks spanning $N$ tasks and $k$ models, let $r_{n,i}$ be the rank of model $i$ on task $n$ (rank 1 $=$ best, with average ranks for ties and the metric direction handled per task). The Friedman
statistic
\begin{equation}
  \chi_F^2 = \frac{12N}{k(k+1)}\Bigl[\sum_{i=1}^{k}\bar{r}_i^2 -
  \tfrac{k(k+1)^2}{4}\Bigr],
  \qquad \bar{r}_i = \tfrac{1}{N}\sum_n r_{n,i},
\end{equation}
rejects the global null of equal mean ranks at level $\alpha$ when it exceeds the appropriate $\chi^2_{k-1}$ quantile. The Nemenyi post-hoc declares two models significantly different when their mean ranks differ by more than the critical difference
\begin{equation}
  CD_\alpha = q_\alpha\sqrt{\frac{k(k+1)}{6N}},
\end{equation}
with $q_\alpha$ the studentised-range quantile for $k$ models at level $\alpha$. The Demšar diagram visualises mean ranks on a number line and joins ``cliques'' of within-$CD$ models with a horizontal bar; this is the standard scalable view when the benchmark spans many tasks.

\begin{figure}[t]
\centering
\resizebox{\linewidth}{!}{%
\begin{tikzpicture}[
    font=\footnotesize,
    >=Latex,
    rankaxis/.style={line width=0.6pt, draw=gnInk},
    ticks/.style={line width=0.45pt, draw=gnInk},
    cdrule/.style={line width=0.55pt, draw=gnInk},
    clique/.style={line width=2.4pt, draw=gnInk, line cap=round},
    callout/.style={font=\scriptsize, text=gnInk, align=left, inner sep=2pt},
    leader/.style={line width=0.35pt, draw=gnRule, dashed,
                   shorten >=1pt, shorten <=1pt},
    arrow/.style={-{Latex[length=1.6mm,width=1.2mm]}, line width=0.5pt,
                  draw=skyblue, shorten >=1.5pt, shorten <=1.5pt},
]
% --- geometry ---------------------------------------------------------
% Diagram zone: x in [0, 10].  Callout column: x >= 10.9.
\def\axL{0}\def\axR{10}\def\axY{0}
\def\rmin{1}\def\rmax{5}
\pgfmathsetmacro{\axW}{\axR-\axL}

% Vertical bands (top -> bottom) avoid every prior overlap:
%   y =  4.20  CD reference label
%   y =  3.80  CD reference bar
%   y =  2.55  staggered model labels (high row)
%   y =  2.10  staggered model labels (low row)
%   y =  0     rank axis
%   y = -0.45  tick labels (1..5)
%   y = -1.05  axis caption
%   y = -1.65  clique bar 1 (A-B)
%   y = -2.10  clique bar 2 (C-D-E)
%   y = -2.85  paperblue brace + caption
\def\cdY{3.80}
\def\cdLabelY{4.20}
\def\labLow{2.10}
\def\labHigh{2.55}
\def\axCapY{-1.05}
\def\cbYa{-1.65}
\def\cbYb{-2.10}
\def\brkY{-2.85}

% --- rank axis --------------------------------------------------------
\draw[rankaxis] (\axL,\axY) -- (\axR,\axY);
\foreach \r in {1,2,3,4,5} {
    \pgfmathsetmacro{\xx}{\axL + (\r-\rmin)/(\rmax-\rmin)*\axW}
    \draw[ticks] (\xx,\axY) -- (\xx,\axY-0.14);
    \node[font=\footnotesize, below=2pt] at (\xx,\axY-0.14) {\r};
}
\node[font=\footnotesize] at (\axL+\axW/2, \axCapY)
    {Mean rank across $N$ tasks (lower is better)};

% --- CD reference scale at the top ------------------------------------
% CD = 1.40 rank units (chosen so A vs C exceeds CD; A--B and C--E
% are within CD).
\def\cdLen{1.40}
\pgfmathsetmacro{\cdPx}{\cdLen/(\rmax-\rmin)*\axW}
\draw[cdrule] (\axL,\cdY) -- (\axL+\cdPx,\cdY);
\draw[cdrule] (\axL,\cdY-0.12) -- (\axL,\cdY+0.12);
\draw[cdrule] (\axL+\cdPx,\cdY-0.12) -- (\axL+\cdPx,\cdY+0.12);
\node[font=\footnotesize] at (\axL+\cdPx/2, \cdLabelY)
    {$CD_{\alpha}=1.40$ rank units \, ($\alpha=0.05$, $k=5$, $N=12$)};

% --- model markers + staggered labels ---------------------------------
\def\rA{1.60}\def\rB{2.10}\def\rC{3.40}\def\rD{3.90}\def\rE{4.30}
\foreach \r/\name/\yy in {%
    \rA/{Model A}/\labLow,
    \rB/{Model B}/\labHigh,
    \rC/{Model C}/\labLow,
    \rD/{Model D}/\labHigh,
    \rE/{Model E}/\labLow
} {
    \pgfmathsetmacro{\xx}{\axL + (\r-\rmin)/(\rmax-\rmin)*\axW}
    \fill[gnInk] (\xx,\axY) circle (1.7pt);
    \draw[leader] (\xx,\axY+0.08) -- (\xx,\yy-0.18);
    \node[font=\scriptsize, align=center, anchor=south]
        at (\xx,\yy-0.14) {\name\\($\r$)};
}

% --- clique bars ------------------------------------------------------
\pgfmathsetmacro{\xClAL}{\axL + (\rA-\rmin)/(\rmax-\rmin)*\axW - 0.18}
\pgfmathsetmacro{\xClAR}{\axL + (\rB-\rmin)/(\rmax-\rmin)*\axW + 0.18}
\pgfmathsetmacro{\xClBL}{\axL + (\rC-\rmin)/(\rmax-\rmin)*\axW - 0.18}
\pgfmathsetmacro{\xClBR}{\axL + (\rE-\rmin)/(\rmax-\rmin)*\axW + 0.18}
\draw[clique] (\xClAL,\cbYa) -- (\xClAR,\cbYa);
\draw[clique] (\xClBL,\cbYb) -- (\xClBR,\cbYb);

% --- significance brace at the bottom ---------------------------------
\pgfmathsetmacro{\xRAx}{\axL + (\rA-\rmin)/(\rmax-\rmin)*\axW}
\pgfmathsetmacro{\xRCx}{\axL + (\rC-\rmin)/(\rmax-\rmin)*\axW}
\pgfmathsetmacro{\xRACmid}{(\xRAx+\xRCx)/2}
\draw[decorate,decoration={brace,amplitude=4pt,mirror},
      line width=0.4pt,draw=paperblue]
    (\xRAx,\brkY) -- (\xRCx,\brkY);
\node[font=\scriptsize, anchor=north, text=paperblue, align=center]
    at (\xRACmid, \brkY-0.18)
    {A\,vs.\,C: rank gap $1.80>CD_\alpha$ $\Rightarrow$ significantly different};

% --- callouts in a clean column on the right --------------------------
% Three numbered callouts vertically stacked, each with a thin grey
% leader pointing to its diagram target.  Pulling them into a column
% prevents collisions with model labels and the axis caption.
\def\coX{10.9}
\node[callout, anchor=north west] (lblCD) at (\coX, \cdLabelY+0.05) {%
    \textbf{(1) Reference scale.}\\
    A pair of models is significantly\\
    different iff their mean ranks\\
    differ by more than $CD_\alpha$.};
\node[callout, anchor=north west] (lblMark) at (\coX, \labHigh+0.05) {%
    \textbf{(2) Model marker.}\\
    Plotted at the model's mean rank\\
    across all $N$ tasks (lower is better).};
\node[callout, anchor=north west] (lblClq) at (\coX, \cbYa+0.20) {%
    \textbf{(3) Clique bar.}\\
    Connects models whose mean-rank\\
    span is below $CD_\alpha$ -- i.e.\\
    not significantly different.};

% Arrows from each callout to its diagram target.
\pgfmathsetmacro{\xCdMid}{\axL + \cdPx/1}
\draw[arrow] (lblCD.west)   -- (\xCdMid, \cdY);
\pgfmathsetmacro{\xMarkA}{\axL + (\rA-\rmin)/(\rmax-\rmin)*\axW}
\draw[arrow] (lblMark.west) -- (\xMarkA, \axY+0.2);
\draw[arrow, shorten >=0pt] (lblClq.west)  -- (\xClBR, \cbYb);

\end{tikzpicture}}
\caption{Anatomy of a cross-category Demšar critical-difference (CD) diagram. \textbf{Top:} the $CD_\alpha$ reference scale, fixed by $k$ and $N$ through $CD_\alpha = q_\alpha\sqrt{k(k+1)/(6N)}$, calibrates which mean-rank gaps are large enough to be significant. \textbf{Middle:} each model is plotted at its mean rank across the $N$ tasks; lower is better. \textbf{Bottom:} clique bars connect contiguous runs of models whose total mean-rank span is below $CD_\alpha$, they are not significantly different from one another at level $\alpha$. Models that share no bar (here A and C) are declared significantly different. The diagram is the standard scalable view when the benchmark spans many tasks; in this paper it appears as Figure 3 with $k=4$ encoders and $N=10$ tasks, where $CD_{0.05}=1.48$ exceeds the largest observed mean-rank gap and all four architectures fall in a single Nemenyi clique.}
\label{fig:cd-anatomy}
\end{figure}
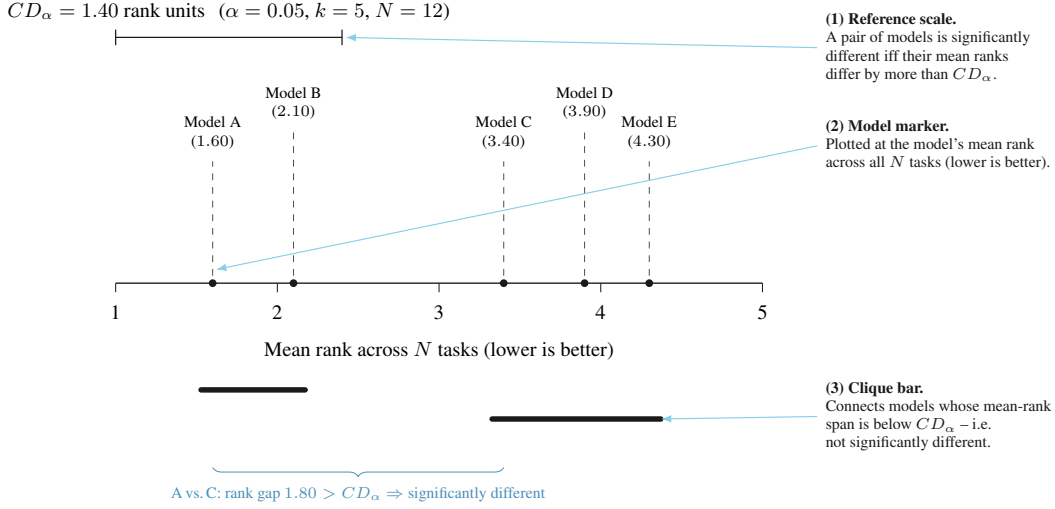

\paragraph{Reading the cross-category Demšar CD diagram.}
The CD diagram answers three questions at once: \emph{(i)} which models rank highest on average across the benchmark, \emph{(ii)} which pairs of models are statistically interchangeable at level
$\alpha$, and \emph{(iii)} what minimum mean-rank gap counts as a real ordering. Figure~\ref{fig:cd-anatomy} labels the visual elements. The reading protocol is:

\begin{enumerate}[leftmargin=1.6em,itemsep=2pt,topsep=2pt]
  \item Locate each model on the horizontal \emph{mean-rank axis}. Lower mean rank is uniformly better by construction, per-task ranks are computed with the metric direction already factored in (higher accuracy/AUC is rank 1, lower MAE is rank 1) and only then averaged across tasks;
  \item Read the \emph{$CD_\alpha$ reference} at the top: any pair of models whose mean ranks differ by more than this length is declared significantly different at level $\alpha$ by the Nemenyi post-hoc. The reference is fixed by $k$ and $N$ through Eq.~6 and is independent of the data;
  \item Inspect the horizontal \emph{clique bars}: each bar joins a contiguous run of models whose mean-rank span is below $CD_\alpha$. Models under a common bar are not significantly different from one another. Models that are not connected by any shared bar are significantly different.
\end{enumerate}

Three caveats. First, $CD_\alpha$ shrinks as $\sqrt{1/N}$, so adding benchmark tasks tightens the threshold proportionally; with $N=10$ the diagram is informative about rank \emph{ordering} but rarely strong enough to certify small gaps as significant. Second, only models present on every task enter the rank table, dropping a single model from one task removes it from the diagram entirely. Third, the diagram makes a statement about average rank, not about absolute metric magnitude: a single-rank improvement on a tiny task
counts as much as a single-rank improvement on a large one.

\section{Bootstrap Confidence Intervals}
\label{app:bootstrap}

The Student's $t$ confidence interval of Appendix~\ref{app:prelim} assumes that per-seed final metrics are approximately normally distributed. For metrics that are bounded away from this assumption, Hits@$K$ on knowledge graphs, MRR, AUC on highly imbalanced edge splits, a percentile bootstrap is more honest. We expose a percentile-bootstrap CI as a drop-in alternative: pass \texttt{method="bootstrap"} to \texttt{report.summary()} or set \texttt{report.ci\_method = "bootstrap"} so that every downstream plot and LaTeX export uses the bootstrap envelope.

\paragraph{Procedure.}
For a fixed (task, model) cell with per-seed final metrics $\{x_s\}_{s=1}^{S}$ we draw $B$ resamples
$\{x^{*}_b\}_{b=1}^{B}$ with replacement, each of size $S$, and compute the resampled means
$\bar{x}^{*}_b = \tfrac{1}{S}\sum_{s} x^{*}_{b,s}$. The $(1-\alpha)$ percentile interval is
$[\,q_{\alpha/2}(\bar{x}^{*}),\ q_{1-\alpha/2}(\bar{x}^{*})\,]$. Because the report API displays a single $\bar{x} \pm h$ figure per cell for symmetry with the $t$-CI form, we report $h = \tfrac{1}{2}\bigl[q_{1-\alpha/2}(\bar{x}^{*}) - q_{\alpha/2}(\bar{x}^{*})\bigr]$, i.e.\ the half-width of a symmetric envelope with the same total width as the percentile interval. We use $B = 10{,}000$ by default and seed the resampling RNG with a configurable \texttt{bootstrap\_seed} so the bootstrap is itself reproducible.

\paragraph{When to prefer it.}
At $S = 10$ seeds the $t$-CI and bootstrap envelopes coincide for near-Gaussian metrics; the bootstrap is meaningfully tighter when the per-seed distribution is skewed (e.g.\ a single seed lands far above or below the others on AUC). The bootstrap also degrades more gracefully than the $t$-CI on metrics with hard ceilings: a sample clustered at $0.99$ test accuracy yields a percentile interval that respects the upper bound, whereas the $t$-CI happily extends past $1.0$.

\section{Custom-Model Templates}
\label{app:templates}

\textsc{GraphNetz} ships five built-in architectures (Section~\ref{sec:design}) behind a small registration surface so user-defined models flow through the same multi-seed, Holm-corrected, CD-diagrammed pipeline. We document the three integration paths below; all three end at the same \texttt{run\_benchmark} call.

\paragraph{Path A: decorator (recommended for libraries).}
Permanent registration at import time. The model becomes visible to every \texttt{run\_benchmark} call by name and the dispatcher skips incompatible (model, task) pairs without further bookkeeping.

\begin{lstlisting}
import torch
from torch_geometric.nn import GCNConv
from graphnetz import register_model, run_benchmark

@register_model(task_type={"node_cls"})
class TinyGCN(torch.nn.Module):
    def __init__(self, in_channels, hidden_channels, out_channels):
        super().__init__()
        self.conv1 = GCNConv(in_channels, hidden_channels)
        self.conv2 = GCNConv(hidden_channels, out_channels)

    def forward(self, data):
        x, ei = data.x, data.edge_index
        return self.conv2(torch.relu(self.conv1(x, ei)), ei)

report = run_benchmark("social", {"TinyGCN": TinyGCN},
                      task_type="node_cls", seeds=(0, 1, 2, 3, 4, 5, 6, 7, 8, 9))
\end{lstlisting}

\paragraph{Path B: class attribute (no decorator dependency).} Equivalent semantics; useful when imports cannot be edited at the top of a third-party file.

\begin{lstlisting}
class AttrGCN(torch.nn.Module):
    task_types = {"node_cls"}
    def __init__(self, in_channels, hidden_channels, out_channels):
        ...
\end{lstlisting}

\paragraph{Path C: inline tuple (one-shot variants).}
Useful for hyperparameter sweeps where each variant needs its own factory. The third slot is a callable \texttt{(in\_channels, hidden\_channels, out\_channels) $\to$ Module}.

\begin{lstlisting}
run_benchmark(
    "social",
    {
        "MyGNN-d0.3": (MyGNN, "node_cls",
                       lambda i, h, o: MyGNN(i, h, o, dropout=0.3)),
        "MyGNN-d0.5": (MyGNN, "node_cls",
                        lambda i, h, o: MyGNN(i, h, o, dropout=0.5)),
    },
    seeds=(0, 1, 2, 3, 4, 5, 6, 7, 8, 9),
)
\end{lstlisting}

\paragraph{Multi-task factory.}
For node-level encoders that should run on \emph{all four} task types without writing the adapter glue, we expose \texttt{\_multi\_task\_factory(cls)}, which wraps the encoder with the appropriate adapter (graph-level pooling head, dot-product link-prediction head, or DGI head) given the dispatched task type.

\begin{lstlisting}
from graphnetz.benchmark import _multi_task_factory, register_model

class MyEncoder(torch.nn.Module):
    """Returns per-node embeddings of shape [N, hidden_channels]."""
    ...

ALL_TASKS = {"node_cls", "graph_cls", "graph_reg", "link_pred"}
register_model(MyEncoder, task_type=ALL_TASKS,
               factory=_multi_task_factory(MyEncoder))
\end{lstlisting}

\section{Custom-Dataset Templates}
\label{app:datasets}

The catalog ships 63 loaders, but \textsc{GraphNetz} is not a closed catalog: any object satisfying the standard PyG contract can be plugged in as a \texttt{Task} and run through the same statistical pipeline.

\paragraph{Path A: ad-hoc \texttt{tasks=} (no global state).}
The simplest form. Wrap an already-loaded dataset with \texttt{task\_from\_dataset} and pass it to \texttt{run\_benchmark} through the \texttt{tasks=} argument; nothing is registered globally, so test code that uses this path leaves the \texttt{BENCHMARK\_TASKS} dictionary untouched.

\begin{lstlisting}
from graphnetz import GCN, run_benchmark, task_from_dataset

ds = my_loader("data/my_dataset")  # any PyG-shaped dataset
task = task_from_dataset("my_dataset", "node_cls", ds, epochs=100)

report = run_benchmark(
    models={"GCN": GCN},
    tasks=[task],
    seeds=(0, 1, 2, 3, 4, 5, 6, 7, 8, 9),
)
\end{lstlisting}

\paragraph{Path B: permanent registration.}
For datasets that should appear in \texttt{iter\_benchmark\_tasks(category=...)} alongside the built-ins. Always pair with \texttt{unregister\_task} in tests so state does not leak between cases.

\begin{lstlisting}
from graphnetz import register_task, task_from_dataset, unregister_task

register_task("my_lab", task_from_dataset(
    "my_assay", "graph_cls", ds, epochs=50))

# ... later ...
unregister_task("my_lab", "my_assay")
\end{lstlisting}

\paragraph{Path C: seed-aware loader for synthetic data.}
For generators rather than fixed datasets, write a loader that takes a \texttt{seed} keyword argument. The dispatcher detects the parameter via \texttt{inspect.signature} and threads the benchmark seed into the loader, so cross-seed variance reflects \emph{both} model initialisation and data resampling instead of only the former.

\begin{lstlisting}
from graphnetz.benchmark import Task

def synthetic_loader(_root, *, seed):
    return MySyntheticDataset(num_graphs=100, seed=seed)

task = Task("synthetic_g100", "graph_cls", synthetic_loader, epochs=20)
run_benchmark(models={"GCN": GCN}, tasks=[task],
              seeds=(0, 1, 2, 3, 4, 5, 6, 7, 8, 9))
\end{lstlisting}

\paragraph{Conventions per task type.}
A custom dataset must satisfy the conventions of the trainer for its declared \texttt{task\_type}: a \texttt{Data} object with \texttt{train\_mask}/\texttt{val\_mask}/\texttt{test\_mask} for \texttt{node\_cls}; a graph-level \texttt{y} for \texttt{graph\_cls} and \texttt{graph\_reg}; an \texttt{edge\_index} suitable for \texttt{RandomLinkSplit} for homogeneous \texttt{link\_pred}; and \texttt{train\_edge\_index}/\texttt{train\_edge\_type}, plus the \texttt{valid\_*}/\texttt{test\_*} counterparts and \texttt{num\_relations}, for relational \texttt{link\_pred}. The runner detects the relational variant from the data attributes and routes it through a DistMult decoder automatically; nothing  else in the call site changes.

\section{Reproducibility and RNG Reseeding}
\label{app:reproducibility}

Algorithm~\ref{alg:bench} reseeds every random number generator the
training code reaches before each \emph{(task, model, seed)} triple:
Python \texttt{random}, NumPy, Torch CPU, and Torch CUDA (when
available). With a fixed software stack -- \texttt{torch >= 2.6},
\texttt{torch-geometric >= 2.6}, the same CUDA driver -- the entire
pipeline is deterministic.

\paragraph{Combinatorial seed threading.}
For loaders whose signature exposes a \texttt{seed} keyword (the
synthetic combinatorial generators in
Section~\ref{sec:design} and any user-defined seed-aware loader from
Appendix~\ref{app:datasets}), the dispatcher passes the
benchmark seed into the loader as well. This is a small but
methodologically important detail: cross-seed variance for
synthetic-data benchmarks then reflects both model initialisation
and data resampling, instead of only the former. For loaders without
a \texttt{seed} keyword the dataset is loaded once and reused
across seeds; the runner caches it to avoid re-downloading.

\paragraph{Software stack used for the headline experiment.}
The numbers in Section~\ref{sec:experiment} were produced with
Python 3.12, \texttt{torch} 2.6.0, \texttt{torch-geometric} 2.6.1,
and \texttt{scipy} 1.13 on a CPU runner. All seeds in
\texttt{(0, 1, 2, 3, 4, 5, 6, 7, 8, 9)} were used; the resulting
pickled \texttt{BenchmarkReport} cache is part of the artefact
bundle so the figures and tables can be regenerated without
retraining.

\section{Broader Impact and Significance}
\label{sec:impact}

\textsc{GraphNetz} does not propose new models or datasets and so has no direct deployment impact. Its intended effect is methodological: lower the activation energy required to report GNN results with confidence intervals and corrected pairwise tests, so that the modal benchmark paper in the field communicates calibrated rather than over-confident evidence. The same pipeline can be applied to any future GNN architecture or dataset added to the catalog, and the methodological apparatus generalizes to any benchmark whose output is a per task, method, and seed metric tensor.

% \newpage
% \input{checklist.tex}

\end{document}